\documentclass[sigconf,nonacm]{acmart}

\usepackage{multirow}
\usepackage{subcaption}
\usepackage{booktabs}
\usepackage{colortbl}
\usepackage{tabularx}
\usepackage{hhline}
\usepackage{appendix}
\usepackage{enumitem}
\usepackage{hyphenat}
\usepackage{balance}

\definecolor{customblue}{HTML}{83b6d4}
\definecolor{customyellow}{HTML}{feb24c}
\definecolor{customred}{HTML}{e34a33}

\AtBeginDocument{%
  }

\begin{document}

\title{Finding Phones Fast: Low-Latency and Scalable Monitoring of Cellular Communications in Sensitive Areas}

\def\name{WaveTag}

\author{Martin Kotuliak}
\affiliation{
  \institution{ETH Zurich}
  \city{Zurich}
  \country{Switzerland}}
\email{martin.kotuliak@inf.ethz.ch}

\author{Simon Erni}
\affiliation{
  \institution{ETH Zurich}
  \city{Zurich}
  \country{Switzerland}}
\email{simon.erni@inf.ethz.ch}

\author{Jakub Polak}
\affiliation{
\institution{Unaffiliated}
\city{Zurich}
\country{Switzerland}
}
\email{polakjakub@icloud.com}

\author{Marc Roeschlin}
\affiliation{
\institution{Unaffiliated}
\city{Zurich}
\country{Switzerland}
}
\email{mar.roe@yahoo.com}

\author{Richard Baker}
\affiliation{
  \institution{University of Oxford}
  \city{Oxford}
  \country{United Kingdom}}
\email{richard.baker@cs.ox.ac.uk}

\author{Ivan Martinovic}
\affiliation{
  \institution{University of Oxford}
  \city{Oxford}
  \country{United Kingdom}}
\email{ivan.martinovic@cs.ox.ac.uk}

\author{Srdjan Čapkun}
\affiliation{
  \institution{ETH Zurich}
  \city{Zurich}
  \country{Switzerland}}
\email{srdjan.capkun@inf.ethz.ch}

\renewcommand{\sectionautorefname}{Section}
\renewcommand{\subsectionautorefname}{Section}
\renewcommand{\subsubsectionautorefname}{Section}

\begin{abstract}

The widespread availability of cellular devices introduces new threat vectors that allow users or attackers to bypass security policies and physical barriers and bring unauthorized devices into sensitive areas. These threats can arise from user non-compliance or deliberate actions aimed at data exfiltration/infiltration via hidden devices, drones, etc. We identify a critical gap in this context: the absence of low-latency systems for high-quality and instantaneous monitoring of cellular transmissions. Such low-latency systems are crucial to allow for timely detection, decision (e.g., geofencing or localization), and disruption of unauthorized communication in sensitive areas. Operator-based monitoring systems, built for purposes such as people counting or tracking, lack real-time capability, require cooperation across multiple operators, and thus are hard to deploy. Operator-independent monitoring approaches proposed in the literature either lack low-latency capabilities or do not scale. 

We propose WaveTag, the first low-latency, operator-independent, and scalable system designed to monitor 5G and LTE connections across all operators prior to any user data transmission. WaveTag consists of several downlink receivers and a distributed network of uplink receivers that measure both downlink protocol information and uplink signal characteristics at multiple locations to gain a detailed spatial image of uplink signals. WaveTag then aggregates the recorded information, processes it, and provides a decision about the connection before the UE completes connection establishment. To evaluate WaveTag, we deployed it in the context of geofencing, where WaveTag was able to determine whether the signals originate from inside or outside of an area within 2.3 ms of the initial base station-to-device message, therefore enabling prompt and targeted suppression of communication before any user data was transmitted. WaveTag achieved 99.66\% geofencing classification accuracy, using only the information collected before connection establishment. Finally, we conduct a real-world uplink measurement evaluation on a commercial 5G SA network.

\end{abstract}

\maketitle

\hyphenation{WhiteRabbit o-pe-ra-tor-in-de-pen-dent in-de-pen-dent}

\section{Introduction}\label{sec:introduction}

In certain scenarios, cellular communication is undesirable or forbidden, demanding strict monitoring and rapid response. Examples include exam halls, where phones are used to cheat~\cite{noauthor_cheating_2019, adams_more_2018, noauthor_scanners_2015}; correctional facilities, where smuggled phones threaten security~\cite{noauthor_french_2025, noauthor_neo-nazi_2024, anderson_contraband_2019, russo_evaluating_2016}; protected airspace, where a drone controlled over a cellular connection~\cite{noauthor_era_2021, dji_enterprise_introducing_2024} can wreak havoc; or confidential voting processes, such as a jury deliberation room~\cite{noauthor_ashley_nodate, zora_real_2012} or even the Vatican conclave~\cite{noauthor_vatican_nodate}, where unauthorized communication could influence the outcomes. Consequently, there is a strong demand for monitoring solutions that can provide detailed real-time information about cellular devices within a sensitive area, fast enough to prevent unauthorized communication. Low latency is crucial, as targeted blocking of a malicious connection requires a decision from the monitoring system before the connection is fully established and user data is transmitted or received~\cite{erni_adaptover_2022, erni_glados_2025, capota_intelligent_2023}.

In addition to being low-latency, such monitoring solutions also need to be scalable (i.e., handle a large number of operators, cells, and UEs in parallel) and integrate with geofencing or localization. This can be challenging, particularly because UEs transmit only a limited number of messages during connection establishment and typically operate in complex, non-line-of-sight environments.

Cellular operators could provide such monitoring solutions~\cite{cherian_lte_2013, margolies_can_2017}. However, several reasons make this impractical:
(1) Accuracy and Cost: Under the current LTE specification~\cite{noauthor_lte_nodate, noauthor_stage_nodate}, operators cannot monitor a user’s connection with the precision required in these situations~\cite{cherian_lte_2013, mahyuddin_overview_2017}, especially in NLoS scenarios involving many indoor locations. Improving precision around sensitive areas would require deploying a much denser base-station infrastructure than exists today.
(2) Latency: No 3GPP-specified solution provides a geofencing result based solely on initial uplink messages; localization happens later in the connection. Without a specification change, an operator aiming to prevent communication from sensitive areas would need to block all communication until geofencing results arrive. %
(3) Feasibility: Every operator in the country would need to deploy this solution and collaborate with sensitive-area administrators. If even one operator refuses or is unable to cooperate, an adversary could bypass the system through that operator. Such nationwide coordination seems infeasible in practice.

On the other hand, state-of-the-art research~\cite{kotuliak_ltrack_2022, oh_enabling_2024, kumar_lte_2014} explores operator\hyp{}independent monitoring but faces practical limitations. (1) None are shown to be low-latency when deployed in a distributed system. (2) They are restricted to a single cell per SDR, whereas realistic deployments involve multiple cells across different operators. (3) They combine downlink and uplink receivers into a single device, limiting their use to locations with good downlink visibility. These challenges make it difficult to collect high-quality signal metrics and integrate them with geofencing systems. (4) They do not support 5G SA.

In this paper, we introduce \name{}, a low-latency 5G and LTE monitoring system independent of cellular operators that can provide information about a cellular connection no later than 2.3 ms after the initial Random Access Response message. \name{} works in parallel on any frequency band across all operators and cells. The signal characteristics of every uplink message are measured by a distributed set of uplink receivers both inside and around the sensitive area, and the measurements are aggregated in a central unit, which can run further analysis steps. Due to its low latency, \name{} provides enough time for other systems to react before any user data reaches the network.

To showcase the capabilities of \name{}, we deployed it in the context of geofencing that combines metrics using a robust machine learning model. After aggregating the signal characteristics, the model classifies the communications as originating from inside or outside of a predefined restricted area. We deployed and tested our system in two different building complexes in urban settings where mobile phones are not allowed. In the two deployments, we installed six and ten uplink receivers, respectively. %
We also developed a platform with multiple UEs continuously re-connecting to the cellular network, in order to obtain a large dataset of connection attempts from inside and outside the building complex. Using \name{}, we achieved 99.7\% geofencing accuracy. %

Finally, we evaluated \name{} on a real-world 5G SA network. First, we validated that \name{} can process multiple 5G cells on the same frequency, as well as measure signal characteristics from UEs connecting to a 100 MHz 5G TDD cell. Then, we evaluated \name{} on a large open field against a commercial smartphone. This evaluation showed that existing LTE features are still valid in the context of 5G SA, and there is only marginal time overhead in the collection of 5G SA signal characteristics, compared to LTE.

In summary, we make the following contributions:
\begin{itemize}%

    \item We design and implement \name{}, the first operator\hyp{}independent, low-latency 5G and LTE monitoring system capable of making a decision about a connection within 2.3 milliseconds of the initial base station-to-device message, before any user data transmission occurs. 
    \item We develop a scalable architecture consisting of one or more downlink receivers and a distributed set of uplink receivers, enabling precise spatial geofencing or localization even in complex, non-line-of-sight environments. 
    \item We present an end-to-end prototype and full-system evaluation, validating the feasibility, scalability, and performance of \name{} in real-world scenarios with more than 100 base stations across four operators and all frequency bands.
    \item We perform scalability, latency, and feature extraction tests in a commercial 5G SA network.
    \item We build a new data collection platform for collecting connection data based on COTS modems and collect a large dataset of \textgreater{}170,000 connections with all the signal features for the initial UL messages. We make this dataset public.
    \item We integrate our system with a machine learning-based geofencing mechanism that achieves 99.7\% classification accuracy using only uplink messages before a connection with the network core is established, enabling early and targeted response to unauthorized communication attempts. %

\end{itemize}

\section{Background}\label{sec:background}

In 5G/LTE cellular networks, users communicate using user equipment (UE) over a wireless channel with base stations (gNodeB for 5G and eNodeB for LTE, we use the umbrella term xNB for both of them). Communication from the xNB towards the UE is the downlink (DL), and the opposite direction is the uplink (UL). Downlink and uplink are separated by frequency---Frequency Division Duplex (FDD)---or time---Time Division Duplex (TDD).

The International Telecommunication Union (ITU) defines allocations of radio spectrum for uplink and downlink communication across a number of frequency bands. In the country of our deployments, five FDD bands are allocated for 5G or LTE ~\cite{3gpp_ts_36101_3gpp_2024} and one extra TDD band for 5G. Operators tend to have frequencies allocated in most of the available bands and set up cells in all of their allocated frequencies. Therefore, any monitoring system must cover all frequency bands allocated to operators simultaneously. 

\subsection{Connection Establishment}

\begin{figure}
    \centering
    \includegraphics[width=\linewidth]{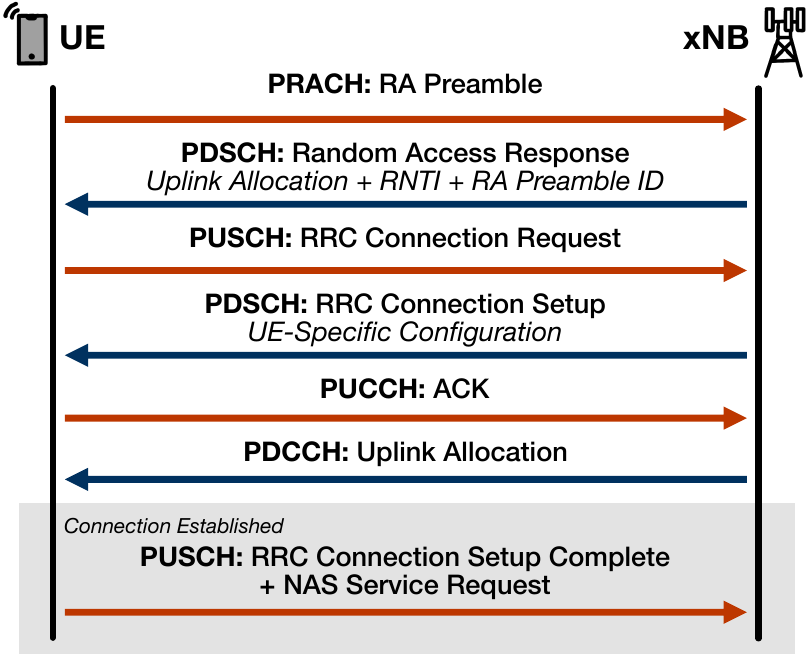}
    \caption{Diagram of a connection establishment procedure between a UE and an xNB.}
    \label{fig:conn_est}
\end{figure}

\autoref{fig:conn_est} shows the initial message exchange between a UE and an xNB during the connection establishment procedure. Initially, the UE scans the spectrum for available xNBs and acquires their configuration broadcast in system information blocks (SIBs). After selecting a suitable cell, e.g., depending on signal conditions and/or operator preferences, the radio connection is initiated with the UE transmitting a Random Access message on the Physical Random Access Channel (PRACH). Upon detecting this message, the xNB responds with a Random Access Response (RAR) message on the Physical Downlink Shared Channel (PDSCH) containing the original PRACH preamble identifier, a unique connection identifier (RNTI), a Timing Advance (TA) command, and a resource allocation specifying time and frequency parameters for the next uplink message of the UE.

This next uplink message, transmitted on the Physical Uplink Shared Channel (PUSCH), is a Radio Resource Control (RRC) Connection Request (RRC Setup Request in 5G). In response to the request, the xNB sends an RRC Connection Setup (RRC Setup in 5G) message on the PDSCH containing dedicated connection configuration parameters for the UE. The UE acknowledges this message on the Physical Uplink Control Channel (PUCCH), and once the xNB sends another uplink resource allocation on the Physical Downlink Control Channel (PDCCH), the UE transmits the RRC Connection Setup Complete and Non-Access Stratum (NAS) Service/Attach Request messages on the PUSCH. Notably, the NAS Service/Attach Request is the first message sent from the UE to the network core, as all prior messages are exchanged solely between the UE and the xNB. We consider the connection to be re-established once the NAS Service Request is delivered, since the following messages are sent encrypted, integrity protected, and can contain user data.

\subsection{Physical Uplink Channels}

On the physical layer, a UE sends a message in a resource grid consisting of OFDM symbols with multiple subcarriers in each symbol. The subcarrier spacing is 15kHz for all the monitored FDD bands, and 30kHz for the 5G TDD band. For 15kHz subcarrier spacing, 14 OFDM symbols form a 1 millisecond long subframe, and 10 subframes form one frame. For 30kHz subcarrier spacing, one subframe consists of 28 OFDM symbols instead. One physical resource block (PRB) spans 12 subcarriers over 7 OFDM symbols. %

\begin{figure}
    \centering
    \includegraphics[width=\linewidth]{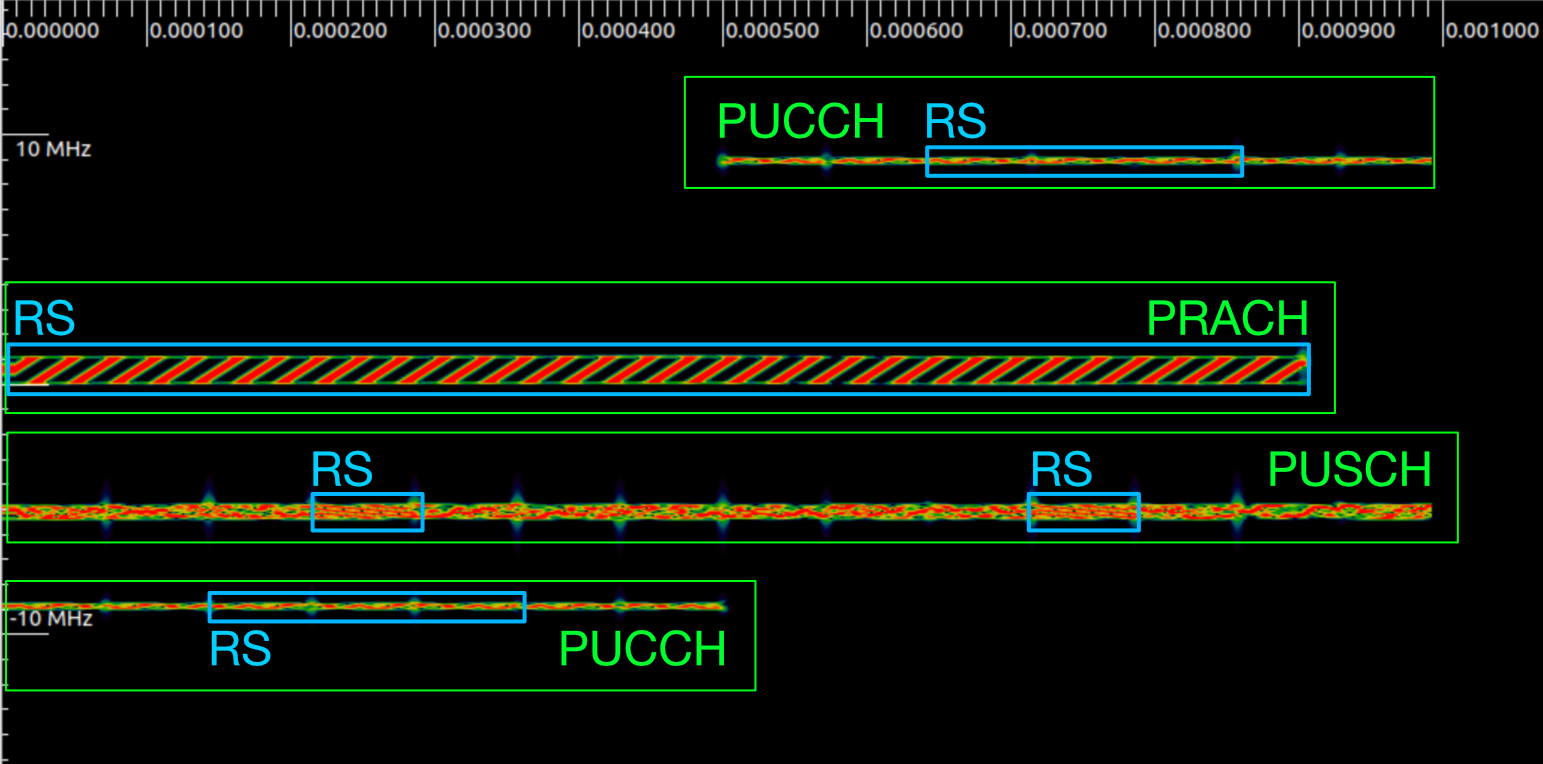}
    \caption{Spectrogram of the LTE uplink for one subframe with all the possible uplink physical channels. The uplink reference signals (RS), which are used for uplink measurements (\autoref{sec:features}), are highlighted in blue. The X-axis represents the time, and in this figure it spans one subframe, lasting 1 millisecond. The Y-axis represents the frequency; in this case, the uplink channel is 20 MHz wide.}
    \label{fig:grid}
\end{figure}

In the uplink, there are three types of physical layer channels: the PUSCH used for transmitting user or control data, the PUCCH used for control information such as ACK/NACK or scheduling requests, and the PRACH used for the initial connection message. All uplink messages in these channels carry a known reference signal alongside the modulated data, which is used for channel estimation, or PRACH detection. Depending on the message, the uplink reference signal spans different OFDM symbols in the resource grid. The reference signal is a Zadoff-Chu sequence, a sequence known for its good autocorrelation properties and constant-amplitude signal. \autoref{fig:grid} illustrates these three types of messages in the spectrum plot and highlights the reference symbols.

Depending on the configuration inside the SIB, a PRACH preamble consists of one or more 839- or 139-symbol-long Zadoff-Chu sequences spanning six PRBs. The UE randomly chooses one of the pre-configured Zadoff-Chu sequences and transmits the preamble in one of the pre-configured PRACH occasions. Upon detection, the xNB replies with a random access response, which contains the preamble index of the chosen sequence. The UE continues the connection establishment only if the preamble index in RAR matches the transmitted PRACH preamble index.

The UE sends a PUSCH message only after it receives an uplink allocation from the base station. The uplink allocation specifies the subframe of the uplink message, the frequency (subcarriers) of the uplink message, and the configuration parameters of the reference symbols within the uplink message, such as which symbols during the PUSCH transmission are reserved for the reference signals. In LTE, 2 out of 14 symbols are allocated to reference signals.

A PUCCH message is sent in pre-configured subframes by the UE or after a downlink transmission as an ACK/NACK. In LTE, PUCCH transmission occupies exactly one PRB on the edge of the cell's spectrum and lasts one subframe. Hopping is often configured for PUCCH, meaning that the first half of the PUCCH is transmitted on the lower edge of the spectrum, and the second half is on the upper edge. It also consists of 14 OFDM symbols; however, typically 6 of them are allocated to reference symbols. For 5G, the PUCCH is similar but with many more configurable parameters.%

\subsection{Physical Downlink Channels}

Similar to the uplink, the physical layer of the downlink also uses OFDMA, with its resource grid split into channels. The PDCCH contains information about the resources for the current downlink transmissions and can also contain uplink allocations for upcoming user transmissions. In contrast, the PDSCH carries cell data, which can include unicast user data, broadcast paging messages, or configuration messages (e.g., SIBs).

\section{System Design}\label{sec:design}

\begin{figure*}
    \centering
    \includegraphics[width=\linewidth]{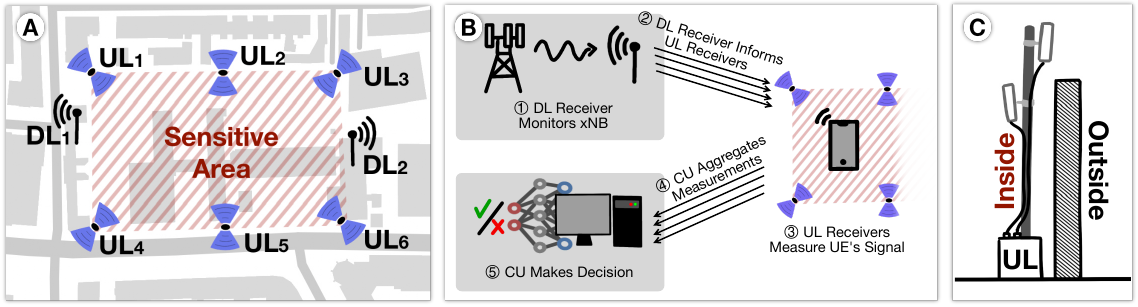}
    \caption{In \textbf{(A)}, we show a sketch of a deployment with two downlink receivers and six uplink receivers positioned around the sensitive area. The purple triangles on the uplink receivers show the azimuth of the two opposite-facing antennas. In \textbf{(B)}, we show how the system components interact. First, the downlink receivers detect a new connection on the downlink and inform the UL receivers about it. The UL receivers then measure the corresponding uplink signals. Finally, all measurements are aggregated in the central unit (CU), which makes a decision about the connection. In \textbf{(C)}, we show how a concrete wall can be used for improved inside/outside separation of a single uplink receiver unit.}
    \label{fig:antenna}
\end{figure*}

\name{} monitors the downlink and uplink of all cells (across multiple operators and frequency bands) visible from the sensitive area. For each UE connection originating within the sensitive area, \name{} captures the required uplink signal features with sufficiently low latency to allow intervention \emph{before} any communication with the operator's core network occurs. 

Accurate spatial inference about a UE connection, such as precise localization or geofencing, typically requires multiple observations of uplink messages. Given the limited number of uplink messages before connection establishment with the core network, \name{} overcomes this constraint by deploying a distributed set of uplink receivers strategically positioned in and around the sensitive area.

In this section, we present the design of \name{}, which addresses the three primary challenges that have remained unaddressed until now: scalable handling of multiple operators and frequency bands, adherence to stringent low-latency constraints, and the ability to support numerous distributed uplink receivers. %

\subsection{System Model and Assumptions}

\name{} consists of one or several downlink receivers and a larger number of geographically distributed uplink receivers. The downlink receivers are synchronized to all xNBs around the target area and monitor their downlink channels for new connections. Once a new connection is detected, the downlink receiver looks for uplink allocations that correspond to the connecting UE. Upon receiving an uplink allocation, the downlink receiver notifies all uplink receivers about the upcoming uplink transmission. All uplink receivers then capture the uplink message and measure its signal characteristics without the need to decode the data contained in the uplink messages. The measurements from all uplink receivers are aggregated in a central unit (CU), where they can be further processed. In our use case shown in \autoref{sec:features}, a model classifies the connection as originating from within or outside the sensitive area. This system model is depicted in \autoref{fig:antenna} in an example deployment.

Due to the requirement for UE detection and decision before full connection establishment with the core network, \name{} collects metrics only for messages sent by the UE before the full connection establishment. This includes at most one PRACH preamble, one RRC Connection Request on PUSCH, and one or two subsequent PUCCH messages. Subsequent PUSCH messages after the RRC Connection Request could already include messages such as a Service Request, reestablishing an existing security context and effectively communicating with the core network.

While we do not impose strict restrictions on the size or shape of the sensitive area that needs protection, our system is intended for sensitive areas that can encompass multiple buildings and span thousands of square meters. We note that those areas are typically separated from the outside by a wall or fence, as this often applies to sensitive perimeters. These facilities can be located in dense urban environments and therefore the separation between inside and outside can be as low as 5 meters, requiring sufficiently precise spatial information for the classification mechanism to work. Furthermore, built-up areas exhibit rich radio frequency environments with several xNBs operated by different cellular network operators, leading to a predominance of non-line-of-sight (NLoS) signal paths between xNB and user equipment.

\subsubsection{Attacker Model}
An \emph{adversary} aims to establish a cellular link from within a sensitive area without being detected. The adversary can use any commercial off-the-shelf (COTS) device that uses a specification-compliant cellular baseband chipset such as smartphones or cellular modules. Additionally, the adversary may use any software on such COTS devices, including running in higher privilege mode (e.g., a rooted phone).
We further assume that the adversary is able to obtain any SIM card (e.g., roaming SIM card) in order to connect to any operator in the area. We also allow the adversary to interfere with the cell selection mechanism by, e.g., forcing the baseband chip to connect to any visible cell, even if further away or with a worse SNR. Finally, we assume the attacker does not use bulky analog radio equipment such as narrow-beam directional antennas, large poles or similar, as any equipment except COTS devices would be spotted during typical entrance inspections in the proposed use cases in the \autoref{sec:introduction}.

\subsection{Downlink Receivers}

The downlink receiver is an SDR-based platform that has the task of monitoring the downlink of cells visible on all bands from the sensitive area. %
To decrease inter-cell interference and enhance decoding rates, we use narrow beam-width directional antennas. To cover all the directions, we plan for four ideally positioned downlink receivers with a direct line of sight to the xNBs. In this way, each xNB in the vicinity is covered by at least one downlink receiver. A downlink receiver with good downlink reception quality continuously remains synchronized with the xNBs and decodes all messages that are broadcast on the downlink. 

After successfully synchronizing in time and frequency to a cell, the downlink receiver decodes the MIB and SIB and applies the cell-specific configurations. These configurations include cell-specific information required for new UE connection detection and uplink signal measurements. Each new UE connection includes a RAR sent by the xNB as a response to PRACH preamble (see \autoref{fig:conn_est}). By monitoring the RA-RNTIs defined in the SIB messages, we can decode all RARs, and thus detect all new connections on the xNB. 

After the downlink receiver decodes the RAR, it immediately forwards the RAR to the uplink receivers, stores the UE-specific RNTI in the set of active RNTIs, and continues monitoring and decoding subsequent messages addressed for the active RNTIs. In case an RRC Connection Setup message is decoded, the downlink receiver applies the enclosed UE-specific configuration to the decoding process for this particular UE.

\subsection{Uplink Receivers}

Uplink receivers, also operating on an SDR-based platform, are connected to two wide-beam directional antennas facing away from each other, as pictured in \autoref{fig:antenna}. This antenna orientation enables uplink receivers to estimate the direction of the signal (see \autoref{sec:features} for more details). Each SDR streams IQ samples of the uplink of one or more cellular frequency bands to an uplink receiver, where it is buffered for at least 20 milliseconds. One of the main advantages of \name{} over existing solutions is that uplink receivers are \textit{decoupled} from downlink receivers. The uplink receivers do not need to monitor downlink frequencies, and therefore they do not need to acquire a sync to the xNBs or decode downlink messages. Instead, they detect and measure uplink signals using the correct uplink allocations and timings received from the downlink receiver.

A key requirement for \name{} is that the downlink and uplink receivers share a common clock. This ensures that the uplink allocations received by the downlink receiver can be mapped to the correct time and frequency to be observed by the uplink receivers.

All relevant downlink control messages, including uplink allocations from RARs, are sent to all uplink receivers over a low-latency communication link. Each uplink receiver then independently looks up the corresponding recorded time and frequency resources for the uplink message in the buffered IQ stream. \name{} then extracts the signal characteristics (see \autoref{sec:features}) based on the reference signals and finally sends them to the central unit. Using this process, uplink receivers detect and measure PRACH, PUSCH, and PUCCH uplink messages in LTE; and PRACH and PUSCH messages in 5G.

We position uplink receivers around the perimeter of the target area. We try to position them such that there is a line of sight to every location where a UE might be present. In case the UEs are inside a building and it is not possible to have a line of sight, we try to position the antennas such that they are on the clearest path out of the building, e.g., having only a window or a single wall in the path. We found this approach to work well when the target area is relatively large and consists of multiple buildings separated from the outside world by a wall or empty space.

\subsection{Central Unit}

Measurements of uplink messages are sent over a low-latency communication link and aggregated at the central unit. Once we receive measurements from every uplink receiver, we can further process the signal metrics for geofencing (or localization) purposes. 

Each uplink message sent by the UE is processed by the central unit as soon as all uplink receivers have submitted their signal characteristics measurements for that message. This can be a heavy-weight process, for example, running inference on a neural network. In a geofencing context, once the system is asked to make a final decision about the connection, the central unit aggregates the decisions from the individual messages into one result. This combination is a fast operation, as we can expect no more than four different message results (e.g., a downlink message might be split into multiple PDSCH messages requiring multiple PUCCH responses). We can therefore perform this operation in real time.

\section{\name{} Implementation}

\begin{figure*}
    \centering
    \includegraphics[width=\linewidth]{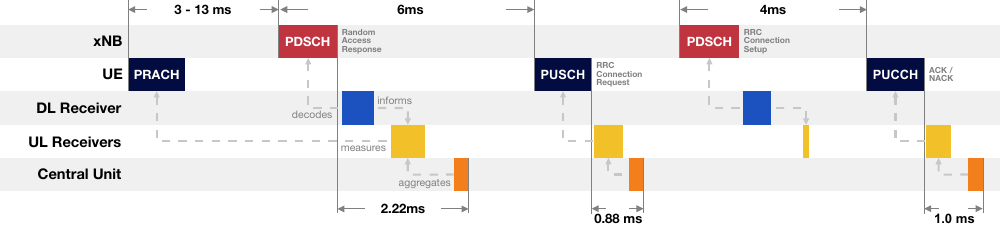}
    \caption{Overview of a typical connection on one of the FDD bands with annotated mean processing latencies for end-to-end measurement of PRACH, PUSCH, and PUCCH transmissions. The downlink receiver decodes the RAR and PDSCH messages. The RAR message both informs about the PRACH message 3-13 ms in the past and the PUSCH message 6 ms in the future. Any PDSCH message triggers a corresponding PUCCH message 4 ms later.}
    \label{fig:latency}
\end{figure*}

In this section, we explain how our system achieves both low latency and real-world scalability, including monitoring multiple cells per operator across all its allocated frequency bands. To this end, we show how the radio units acquire and process IQ samples and how the sub-components communicate with each other.

As an overview, in \autoref{fig:latency}, we show the messages exchanged between the UE and the xNB, the actions of our system's subcomponents in response to these messages, and their associated mean latencies.

\subsection{Downlink Monitoring}

While downlink decoding plays a role in this process, it is not the primary contribution of this paper and is therefore not discussed in detail. Several downlink receiver implementations already exist \cite{erni_adaptover_2022, erni_glados_2025, falkenberg_falcon_2019, bui_owl_2016, ross_towards_2023}. We build our downlink sniffer on top of \cite{erni_glados_2025} and refer the reader to that work for a detailed description of the receiver design and implementation.

\subsection{Uplink Processing}\label{sec:impl:ul_rec}

To extract the metrics of the uplink messages reliably, the uplink and downlink receivers must be synchronized to the same time source with a low absolute pairwise offset. In our deployments, we have created a clock distribution tree that spans across all radio units. At the root, there is a high-precision clock synchronization device, generating highly stable 10 MHz and 1 PPS clock signals, disciplined by GPS. This signal is then distributed with the WhiteRabbit protocol~\cite{serrano_white_2013} to all radio units.

Classically~\cite{kotuliak_ltrack_2022, hoang_ltesniffer_2023}, uplink monitoring of an LTE cell is achieved by tuning an SDR receiver to the center frequency of the uplink of the cell and collecting 30.72 million IQ samples per second. However, in each of the frequency bands, there are three to four operators, each transmitting on different frequencies. Using the classical approach, we would require up to 24 SDRs in each uplink receiver to monitor the entire uplink spectrum. Such a system would be extremely expensive and difficult to maintain.

In our solution, we use SDRs with a sampling rate of at least 122.88 Msps, such that they have enough bandwidth to receive a whole FDD frequency band. For the 5G TDD band with cells having as much as 100 MHz bandwidth, multiple SDRs are needed. To ensure that the DC component does not disturb the uplink frequencies, we set the center frequency of our SDR between two operators allocated within a band, as close to the middle of the frequency band as possible. An uplink receiver continuously receives IQ samples from all SDRs and writes them into circular buffers. We also keep the timestamp of the last written sample. Using this timestamp, we can compute the absolute time of each IQ sample in the buffer.

For each monitored cell, we spawn a new cell object, identified by its technology (LTE or 5G), center frequency, and Physical Cell Identity (PCI). When a cell object receives an uplink allocation from the downlink receiver, it is pushed into an event queue with the corresponding uplink transmission time. Each cell object continuously checks in the circular buffer if it has enough samples to process the earliest uplink allocation. Once samples for the uplink allocation are fetched from the circular buffer, our software extracts the uplink signal features. 

During the feature extraction, \name{} first shifts the copied IQ samples in frequency such that the cell's uplink band is centered in the middle of the received bandwidth. Afterwards, it computes an FFT of size \textit{sampling rate} / \textit{subcarrier spacing} for each OFDM symbol containing a known reference signal. This FFT operation gives us modulated values at each individual subcarrier. Then, only the subcarriers defined in the uplink allocation are further processed, i.e., we compute the circular correlation of the received reference signal with a known reference signal (see \autoref{sec:features} for more details). Given that the number of allocated subcarriers is generally very low for the initial messages, this circular correlation is quick.  %

This processing is executed independently in parallel for each of the two radio ports. In our real-world deployment on the frequency band with the most activity, such a measurement was done on average every 5 milliseconds across all cells.
Once the measurement is completed for both ports, the uplink receiver transmits the signal characteristics to the central unit. %

\subsection{Message Exchange}

The downlink receiver, the uplink receivers, and the central unit continuously communicate with each other over a low-latency link. They are connected with fiber links in a star architecture with a switch in the middle. Given the relatively small number of devices, the overhead of such a network topology is negligible. In \autoref{sec:eval:benchmark}, we evaluate how long a message exchange takes.

We use the ZMQ protocol \cite{noauthor_zeromq_nodate} as it is a low-latency protocol with minimal overhead. We utilize its PUB/SUB pattern, where the transmitters are publishers and the receivers of messages subscribe to the publishers. Essentially, publishers broadcast messages and they do not have any information on the state of the subscribers. However, this also means that if a subscriber goes offline, the publisher does not know about it. In our system, we expect relatively low component downtime, and if any components shut down, the distributed system will be minimally affected. Therefore, a PUB/SUB pattern is a good choice for the minimal overhead it brings. We run all messaging services on a separate thread.

\subsection{Central Unit}

The measurements of the individual messages are sent over ZMQ to a central unit. When the first measurement for a message arrives, we start a timer with a timeout of 2 milliseconds. Once all uplink receivers have delivered the measurements, or when the timer runs out, we process the signal measurements for this uplink message as outlined in the next section. Therefore, even if an uplink receiver malfunctions, it does not have a big impact on the overall latency.

\section{Low-latency Geofencing}\label{sec:features}

The main challenge of precise geofencing is classifying a connection’s origin---i.e., determining whether it is inside or outside a sensitive area---based on as few uplink messages as possible. To achieve this, \name{} uses a combination of relative signal strength and signal quality features. These features are collected from multiple uplink receivers, each equipped with two directional antennas facing in opposite directions. Finally, a classification model running on the central unit infers whether the connection originates inside or outside the perimeter. %

\subsection{Uplink Signal Features}

To measure the uplink signal features, we rely on the PRACH preamble or reference signals sent in PUSCH (and PUCCH in LTE). Since these signals are known from the cell configuration or user-specific configurations, our receivers can efficiently compute the circular cross-correlation of the received signal with the reference, without having to decode the actual messages. The OFDM symbols that contain the data are essentially ignored. Finally, the OFDM Cyclic Prefix allows the cross-correlation to be calculated even if the signal is received with a time offset. %
From the circular cross-correlation, we extract signal power (peak) and signal quality characteristics (peak to average power ratio). In \autoref{sec:features_appendix}, we give more details on what exact features we collect. %

The radio units of our uplink receivers are equipped with two receiving ports. Each port is connected to a separate directional antenna, oriented in opposite directions. Intuitively, if a signal is significantly stronger on one antenna, the signal must originate from the direction of that antenna's azimuth. On the other hand, if the signal power on both antennas is equal, the direction is perpendicular to the azimuth of both antennas. 

\begin{figure}[h]
    \centering
    \includegraphics[width=\linewidth]{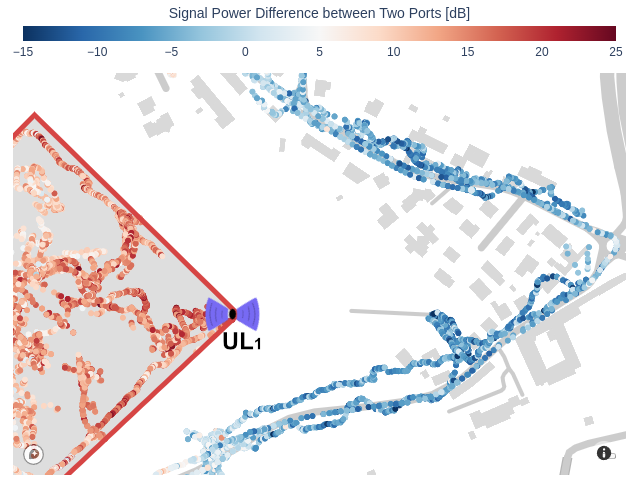}
    \caption{Measured signal power difference between two ports of a single uplink receiver in a real-world deployment. The color shows the measured signal power difference of an uplink signal transmitted from the corresponding point on the map.}
    \label{fig:aoa_sep}
\end{figure}

We can further adapt the direction-finding approach of the two ports to focus on classifying signals as either inside or outside the target area. The most straightforward method involves placing uplink receivers along the border of the target area, with one antenna positioned as an "inside" antenna and the other as an "outside" antenna. Often, the target area is separated from the outside by a boundary, such as a concrete wall or another physical feature. While a boundary such as a wall is not strictly required, it significantly enhances separation when available. If it is not feasible to position antennas on different sides of the boundary, both antennas can be placed on the same side. In such cases, one antenna is positioned above the wall and the other below, ensuring that at least one benefits from the wall’s attenuation effects. This flexible approach enables the system to adapt to various boundary conditions. An example of this antenna setup is shown in \autoref{fig:antenna}. To showcase this separation in the real world, we plot in \autoref{fig:aoa_sep} the relative signal power difference between the two ports of one radio unit for uplink messages collected in one of our deployments.

\subsection{Geofencing Model}\label{sec:impl:model}

Our classification model is a fully connected neural network that uses relative features as input. We first train the model offline with Python and pyTorch~\cite{noauthor_pytorch_nodate}. We use a simple architecture to prevent overfitting and ensure fast inference time. The architecture consists of BatchNorm, Dropout and two fully connected layers followed by a ReLU activation function. Finally, we pass the output neuron's result through a sigmoid function to produce the classification decision. We use binary cross-entropy as a loss function and optimize the model using an Adam optimizer. We then train the model in batches and stop training once the loss no longer improves across several epochs. Finally, we perform a hyperparameter search for each deployment separately to ensure an optimal solution. %

For all connections, we get between 2 and 4 messages before having to make a decision on whether the phone is inside or outside. Thus, we need to combine the decision results (a real number between 0 and 1) of the individual messages. To this end, we train a logistic regression model~\cite{noauthor_logisticregression_nodate}, which weights each classification result based on its message type. If we have not received a given message type, we replace its classification result with 0.5. If we have multiple classification decision results of the same message type, we take their mean value.

\subsection{Data Collection}

For any ML model, the most important factors are the quality of the training data and the number of data points. Generating uplink transmissions with a typical modern smartphone is easily possible, for example, by switching airplane mode on and off. However, there is normally no access to low-level base-band messages from the RRC and NAS layer. We therefore employ a set of cellular COTS modems. These modems have the same functionalities as modern smartphones in terms of cellular communication, and we can use these modems and smartphones interchangeably. However, they also provide low-level debugging features and logs. For example, we can force the modem to connect to a specific frequency band or even a particular cell. It also provides the connection identifier (RNTI), which we use to match the recorded signal features at the central unit with the modem’s recorded location. This gives us a robust way to collect a large amount of labeled training data across multiple operators, bands, and base stations.

To improve the efficiency of the data collection, we fitted a backpack with multiple such modems that connect to the network almost every second. We can then walk a route and easily gather large amounts of uplink signal features from any location. Such an annotated dataset is crucial for training and testing of the model. We make this data available online\footnote{\url{https://doi.org/10.3929/ethz-c-000794381}} to support further research.

\section{Evaluation}\label{sec:eval}

To evaluate the performance of our system, we conducted tests in two different deployments with two different SDR platforms. The first deployment was used as a feasibility study with COTS SDRs and was limited to a single frequency band at a time. The second deployment is our full system, with hardware that is capable of monitoring all frequency bands from all operators in the country simultaneously. For both deployments, we had the necessary approvals from the relevant institutions to conduct the tests and we tested the geofencing performance against our own devices. At the time of these tests, only LTE and 5G-NSA networks were available at these locations. To show that our system is 5G SA capable, we performed a set of tests on a commercial 5G SA network. We confirm that key features---latency, feature extraction, and ability to track multiple cells---still hold for 5G SA.

\subsection{Latency}\label{sec:eval:benchmark}

In this section, we evaluate the latency of individual processing steps and also the end-to-end (E2E) latency of our system for various types of messages. Both individual steps and E2E latencies are shown in \autoref{fig:latency}. We use the Perfetto~\cite{google_googleperfetto_2024} and Tracy~\cite{taudul_wolfpldtracy_2025} libraries to measure the processing times of individual components during \name{}'s operation. The exception is the measurement for the ZMQ message round-trip times, for which we built our own measurement tool.

\begin{table}
    \centering
    \begin{tabular}{lrrr}
    \toprule
     \textbf{Operation} &       \textbf{Mean} & \textbf{StdDev} &  \textbf{Count}\\
\midrule
         Downlink Decoding & 540 $\mu s$ &155 $\mu s$ & 47\\
         Message Round-Trip & 75 $\mu s$ &2 $\mu s$ & 1000\\
         LTE PRACH Measurement & 587 $\mu s$ &147 $\mu s$ & 2027\\
         LTE PUSCH Measurement & 505 $\mu s$ &129 $\mu s$ & 2024\\ 
         LTE PUCCH Measurement & 423 $\mu s$ &145 $\mu s$ & 3534\\
         Model Inference & 254 $\mu$s &78 $\mu s$ & 147\\ \midrule
         \textbf{LTE E2E PRACH} & \textbf{2223 $\mu s$} &\textbf{230 $\mu s$} & \textbf{48} \\
         \textbf{LTE E2E PUSCH} & \textbf{877 $\mu s$} &\textbf{128 $\mu s$} & \textbf{26}\\
         \textbf{LTE E2E PUCCH} & \textbf{1028 $\mu s$} &\textbf{155 $\mu s$}& \textbf{13} \\ \midrule
         5G PRACH Measurement & 479 $\mu s$ &51 $\mu s$ & 140\\
         5G PUSCH Measurement & 602 $\mu s$ &64 $\mu s$ & 140\\
         \bottomrule
    \end{tabular}
    \caption{Latency measurements of both individual parts of the system and the overall E2E latency from when the message is sent by the UE until a decision is available.}
    \label{tab:bench}
\end{table}

In \autoref{tab:bench}, we show the measurement results of the individual parts of the system, as well as the E2E latencies for various message types. For the E2E PRACH measurement, we measure the elapsed time from the reception of the RAR by the downlink receiver to the completion of the inference of the geofencing model. This time encapsulates the downlink decoding of the RAR, the message exchange between the downlink receiver and the uplink receiver, the PRACH measurement by the uplink receiver, the message exchange between the uplink receiver and the central unit, and finally the model inference. For the E2E PUSCH and E2E PUCCH measurements, we measure the time from the reception of the uplink message to the inference result. This measurement does not include downlink decoding and initial message exchange, as \name{} knows about the uplink message already 4 subframes prior to the uplink transmission.  

In summary, we need less than 2.3 ms after the RAR is transmitted to know the result for the corresponding PRACH. For a PUSCH message, the classification is ready on average 0.88 ms after the uplink signal is received. Finally, a PUCCH message is classified on average 1.0 ms after the uplink signal is finished transmitting. The latency of \name{} does not suffer when measuring features in 5G SA compared to LTE. From measurements done on the TDD band while monitoring a commercial 100 MHz cell, we can see a slight improvement of 108$\mu$s for PRACH measurement, and a slight slowdown of 97$\mu$s for PUSCH measurement.

\subsection{Scalability}\label{sec:scalability}

In the second deployment, we observed that \name{} was monitoring more than 100 cells at the same time. Given that we monitor 5 frequency bands, a single SDR can monitor the uplink of more than 20 cells. From our data set, we know that we observed our test phones connecting to 62 of those cells across all bands.

Overall, in the second deployment, we measured 1,172,772 uplink signal messages on 10 uplink receivers, with 620 (less than 0.01\%) not received by the CU.%

\name{} is capable of monitoring multiple cells on the same frequency in parallel on a single SDR port for 5G SA. We evaluate this claim by setting up \name{} on a COTS SDR platform and conducting a 1-hour-long monitoring campaign on two real-world cells in the vicinity. The two cells have a bandwidth of 15 MHz and are operating on the same frequency in an FDD band. Using this setup, we observed 9,106 connections in total, with 8,230 on the first cell and 876 on the second cell. For all connections, we successfully measured both the PRACH and the PUSCH messages. 

Moreover, as discussed in \autoref{sec:impl:ul_rec}, our system is capable of monitoring an entire 100 MHz 5G SA cell. Similarly to before, we ran \name{} on a 100 MHz real-world cell on a TDD band in the vicinity for 1 hour. In this time span, we detected 1,152 connections and we failed to measure only one PRACH for those connections. For PUSCH, we did not observe any failures. 

\subsection{Geofencing}
\subsubsection{First Deployment}

\begin{figure}[ht]
    \centering
    \includegraphics[width=1\linewidth]{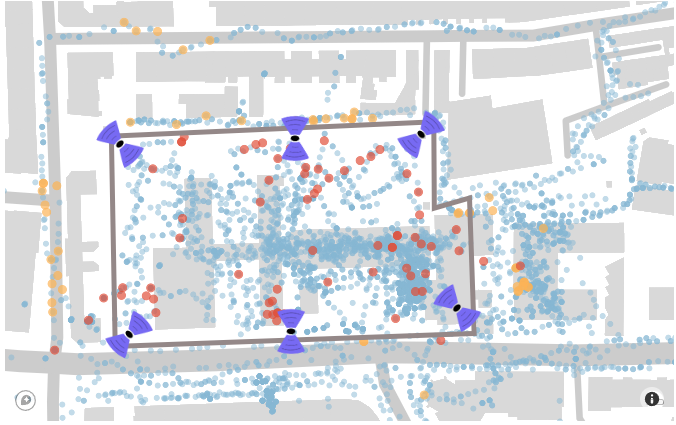}
    \caption{First deployment in an urban area showcasing six uplink receivers (the two triangles on each uplink receiver show the azimuth of their antennas) positioned around the sensitive area. Blue dots represent correct geofencing (inside/outside), with orange dots showing false positives and red dots representing false negatives.}
    \label{fig:mlh}
\end{figure}

The initial feasibility deployment, shown in \autoref{fig:mlh}, was set in the middle of the urban area. The area is approximately 100 $\times$ 200 meters and consists of a single multi-story building with many wings, and many outside areas/courtyards around the building. Part of the sensitive area is divided from the outside world by a concrete wall, but on the right side it is separated from buildings with just a road. The separation between the "inside" and "outside" classes is as little as 3 meters in some areas.

We equipped the site with one downlink receiver placed on the roof of the building and six uplink receiver units positioned around the sensitive area. As radio units, we used COTS SDRs: for the downlink receiver, we used a USRP B210 and for uplink receivers, we used 2 $\times$ USRP X310, 2 $\times$ USRP N310, and 2 $\times$ USRP X410. All SDRs were connected to a central unit using fiber optic cables.

We collected samples over 5 days using our custom data collection rig. We collected training data for 244,804 uplink messages, spanning 43,157 connections. In this set, we have 63.3\% inside and 36.7\% outside connections. To acquire the test set, we collected data on a separate day to ensure that we were not overfitting on specific environmental conditions. The test set consists of 29,809 uplink messages spanning 6,198 connections. 58\% of those are inside connections and 42\% outside connections. Finally, the geofencing model achieves an accuracy of 95.78\% on individual messages from the test set.

\begin{table}[ht]
    \centering
   \begin{tabular}{ll|c|c|l@{}l}
\multicolumn{2}{c}{}&\multicolumn{2}{c}{\textbf{Actual}}&\multicolumn{2}{c}{}\\
\multicolumn{2}{c}{}& \multicolumn{1}{c}{\cellcolor{gray!10} Inside} &\multicolumn{1}{c}{\cellcolor{gray!10}Outside}&Acc & = 97.42\\
\hhline{|~~|--|}
\multirow{2}{*}{\textbf{Predicted}}& \cellcolor{gray!10}Inside & \cellcolor{customblue!20}3503 & \cellcolor{customyellow!20}68 & FPR &= 2.61\%\\
\hhline{|~~|--|}
& \cellcolor{gray!10}Outside & \cellcolor{customred!20}92 & \cellcolor{customblue!20}2535 &FNR\:&= 2.56\%\\
\hhline{|~~|--|}
\end{tabular}
    \caption{Geofencing accuracy results for test set from the first deployment.}
    \label{tab:first_deployment}
\end{table}

A logistic regression model, trained on the training data set, estimates the weight of each type of message in the final ensemble. The PRACH message has a weight of 0.36, PUSCH 0.37, and PUCCH 0.27. As reported in \autoref{tab:first_deployment}, when we run the logistic regression model on the test set, we achieve an accuracy of 97.42\%. From \autoref{fig:mlh}, we see that the model mostly misclassified messages close to the border of the sensitive area, which is expected.

\subsubsection{Second Deployment}

\begin{table}[]
    \centering
   \begin{tabular}{ll|c|c|l@{}l}
\multicolumn{2}{c}{}&\multicolumn{2}{c}{\textbf{Actual}}&\multicolumn{2}{c}{}\\
\multicolumn{2}{c}{}&\multicolumn{1}{c}{\cellcolor{gray!10}Inside}&\multicolumn{1}{c}{\cellcolor{gray!10}Outside}&Acc & = 99.66\%\\
\hhline{|~~|--|}
\multirow{2}{*}{\textbf{Predicted}}& \cellcolor{gray!10} Inside & \cellcolor{customblue!20}10442 & \cellcolor{customyellow!20}30 & FPR &= 0.28\%\\
\hhline{|~~|--|}
& \cellcolor{gray!10} Outside & \cellcolor{customred!20}43 & \cellcolor{customblue!20}10796 &FNR\:&= 0.41\%\\
\hhline{|~~|--|}
\end{tabular}
    \caption{Geofencing accuracy results for test set from the second deployment.}
    \label{tab:second_deployment}
\end{table}

After the successful initial deployment, we deployed our system in a second, larger location. Here, we monitor an area of around 250m $\times$ 250m. The sensitive area consists of multiple buildings with outdoor locations around and in between the buildings. The sensitive area is located on the outskirts of an urban area with neighboring buildings on two sides. We deployed ten prototype uplink receivers based on the custom-built SDR.  

We again collected measurements over multiple days using our custom data collection rig based on LTE COTS modems. For the training set, we collected 486,850 measurements for 103,564 connections over 6 days. The inside/outside split is 52.3\%/47.7\%. We collected the test data set on a separate day with 21,311 connections.

We trained a geofencing model as a fully connected neural network model, and furthermore, we trained a logistic regression that aggregates results over multiple messages. The estimated weights from the logistic regression are: 0.39 for PRACH, 0.38 for PUSCH and 0.23 for PUCCH, approximately matching the weights from the first deployment. The accuracy of the neural network model is 99.02\% on individual messages. The ensemble model then increases the accuracy for the entire connection to 99.66\%. In \autoref{tab:second_deployment}, we report more detailed results for the final geofencing model. Throughout the deployment, we also conducted regular spot-checks with a large number of commercial phones. We did not observe any degradation in performance for any particular model. In \autoref{sec:robust}, we also show how robust our model is under various input constraints.

\subsection{5G SA Feasibility Study}

\begin{figure}[ht]
    \centering
    \includegraphics[width=1\linewidth]{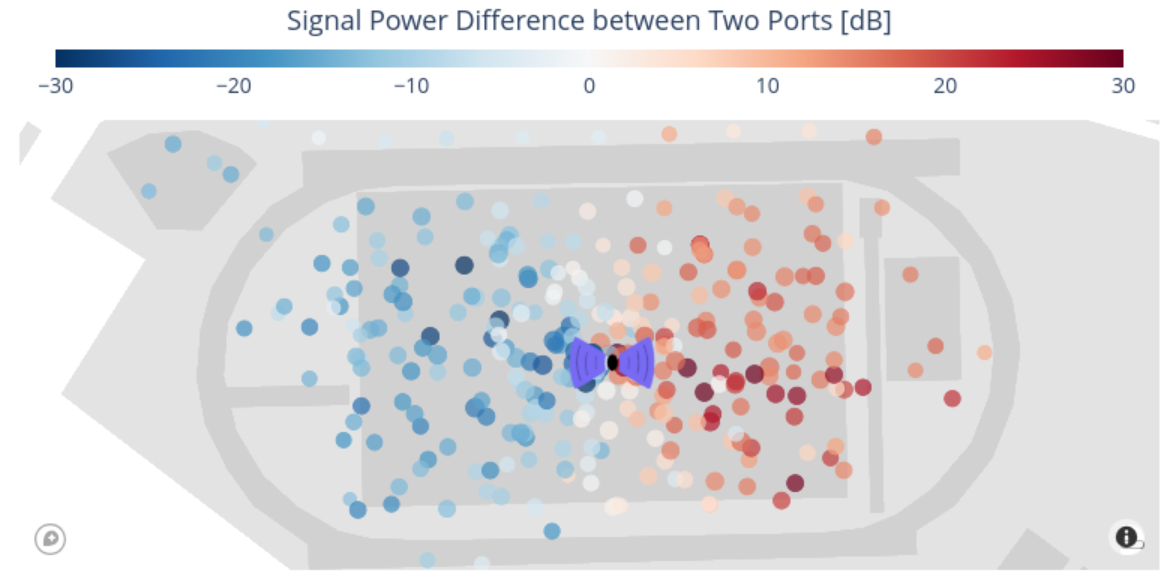}
    \caption{Test of the feature extraction from the 5G PHY layer performed on a commercial 5G SA network. Each point on the map represents a test location. Its color shows the relative signal power measured on two ports of the uplink receiver.}
    \label{fig:NR}
\end{figure}

To evaluate the 5G SA capabilities of \name{}, we set up an uplink receiver running on a USRP X410 right in the middle of a large open field. We used a setup with two directional antennas pointing away from each other. We then conducted an extensive walk around the field with a smartphone (OnePlus 10 Pro 5G) connecting to the commercial 5G SA network. For each connection request, our system measured the features for the PRACH and the first PUSCH message and stored them in a database. Finally, to show our uplink receiver's opposite-facing antenna setup, we plot the relative signal power between the two ports on a map in \autoref{fig:NR}.

This experiment confirms that feature extraction also works reliably in 5G SA. Even at the farthest measurement points from the uplink receiver (\textasciitilde100 m), we observed an SNR exceeding 20 dB. In addition, \autoref{fig:NR} shows a clear spatial separation between measurement points on the left and right sides of the field. This indicates that the relative signal strength measured at the two antenna ports correlates with the direction of the uplink transmission. Given the strong similarities between the LTE and 5G PHY layers, together with the successful field evaluation, we conclude that if 5G SA were deployed at one of our sites, \name{} would also be able to reliably geofence 5G SA connections.

\section{Related Work}

Other operator-independent monitoring systems that rely on uplink measurements include \cite{kotuliak_ltrack_2022, oh_enabling_2024, kumar_lte_2014}. They couple uplink and downlink processing on a single host using one (or more) SDRs, which leads to two drawbacks: (1) \textit{Colocation of downlink and uplink sniffers}, forcing placement where good downlink reception is available and preventing flexible uplink deployment (e.g., indoors or in NLoS environments); and (2) a \textit{one-to-one mapping} between uplink and downlink receivers, increasing hardware and processing requirements. In contrast, our system decouples uplink and downlink receivers, enabling independent placement and improved scalability, as one downlink receiver can serve multiple uplink receivers.

In our deployment, we observed the activity of our phones on 62 different cells in all frequency bands. \cite{kotuliak_ltrack_2022, oh_enabling_2024, kumar_lte_2014} lack the ability to continuously monitor multiple cells simultaneously on a single receiver. As a result, monitoring these 62 cells would require a separate pair of uplink and downlink receivers for each cell, multiplied by the number of distributed locations in a deployment. \name{} processes the uplink signals of the cells in a single frequency band using just one SDR. To put it into perspective, our system in the second deployment could be implemented with as little as 38 COTS SDRs, compared to more than 620 SDRs required by \cite{oh_enabling_2024}.

The target use cases also differ. Prior systems focus on accurate localization over an active connection (e.g., looking for an uncooperative device from a moving vehicle). Because of their use cases, they can rely on a large number of PUSCH observations, rotational antennas, or even active power-boosting attacks~\cite{oh_enabling_2024}. As a result, localization in \cite{oh_enabling_2024} takes around 2 minutes, \cite{kotuliak_ltrack_2022} measures ToA values of all messages from the whole connection, and \cite{kumar_lte_2014} requires offline processing of recorded samples. Even though they achieve their respective tasks, these points show that they are unsuitable for low-latency geofencing. \name{} compensates for the low number of observations with a higher number of uplink receivers, while also measuring PUCCH and PRACH messages.%

Finally, all these works lack the capability to monitor 5G SA. The only existing 5G SA uplink sniffer, Sni5gect \cite{luo_sni5gect_2025}, is built to decode uplink (PUSCH) messages on just one xNB. However, it is not capable of extracting signal features crucial for geofencing and its monitoring capabilities were only evaluated in a 20m range on a lab base station. Even at this small distance, the decoding rate was as low as 55\% for one of the phones. On the other hand, in 5G, our \name{} extracts important signal metrics not only for PUSCH but also for PRACH. Our system and its low-latency signal extraction capabilities were tested on a commercial operator’s network. Even at 100m distances, it consistently measured uplink signals with more than 20 dB SNR. We also showed \name{}’s capability to monitor multiple base stations on a single SDR.

In this paper, we do not try to compare our geofencing accuracy with previous localization techniques. These techniques use TDoA, AoA, or RSSI principles~\cite{hou_efficient_2018, jiang_alrd_2013, he_hybrid_2020, radnosrati_localization_2020} to achieve a sufficiently high localization accuracy (localization being a harder problem to solve than geofencing) but also require expensive optimization algorithms. However, \name{} is extensible and could integrate new localization algorithms for various use cases. To this end, we are open-sourcing our datasets and invite researchers to develop and test different algorithms in a real-world setting, which could then be ported to our platform.

\section{Conclusion}

In this paper, we presented \name{}, a system designed to reliably monitor the uplink of all LTE and 5G connections within sensitive areas while addressing key challenges in latency and scalability. Unlike other systems, \name{} relies only on the initial few messages to make a geofencing decision, enabling low-latency operation critical for rapid countermeasures. Through multiple evaluations on a commercial network, we show that \name{} is the first system in its category capable of monitoring 5G SA. Our system achieves high geofencing accuracy, even in non-line-of-sight environments, as demonstrated through evaluations in two separate deployments using different hardware platforms. In the second, full-scale deployment, \name{} achieved a geofencing accuracy of 99.66\%.

\begin{acks}
The authors would like to thank Patrick Leu for valuable discussions regarding the system design and for insightful suggestions, including the concept of using physical walls to increase the effect of opposing-facing antennas, which helped shape the final implementation.
\end{acks}

\newpage
\balance

\appendix

\begin{appendices}

\begin{figure*}[h]
    \centering
    \begin{subfigure}[t]{0.33\textwidth}
        \centering
        \includegraphics[width=\textwidth]{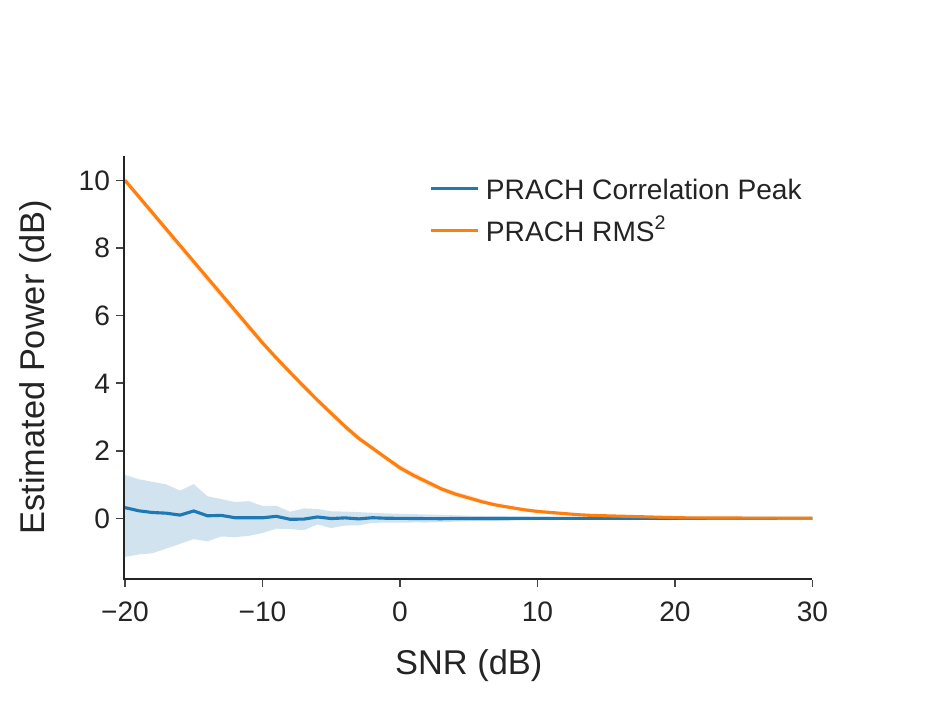} %
        \caption{\parbox{0.9\linewidth}{Estimation of the signal power of a 0 dB signal using two different techniques.}}
        \label{fig:features:sig_power}
    \end{subfigure}
    \hfill
    \begin{subfigure}[t]{0.33\textwidth}
        \centering
        \includegraphics[width=\textwidth]{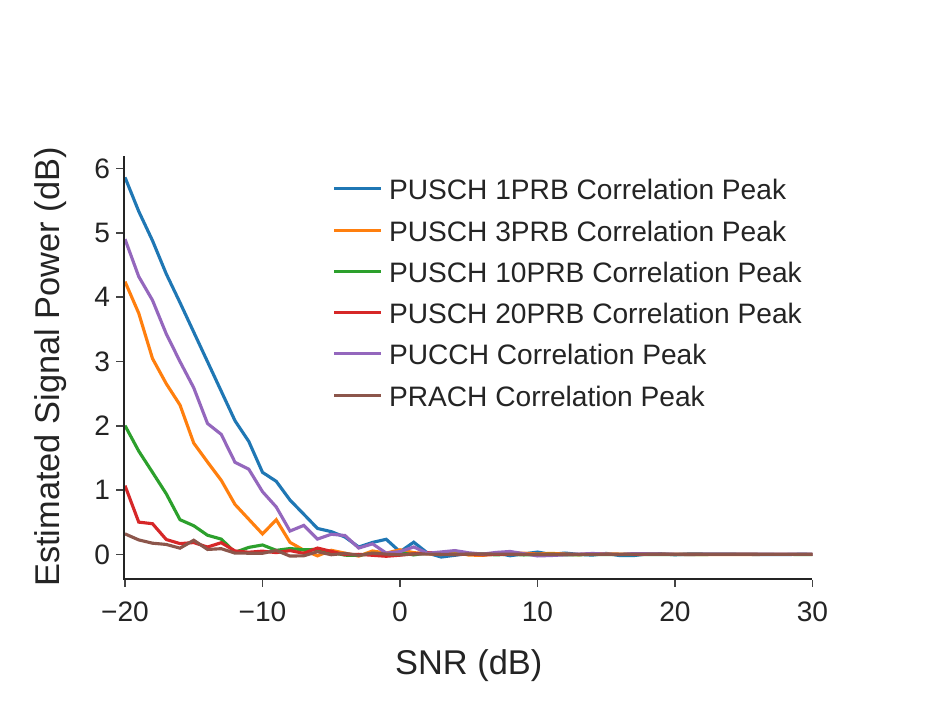} %
        \caption{\parbox{0.9\linewidth}{Estimation of the signal power of various messages sent with the same 0 dB power with the Correlation Peak technique.}}
        \label{fig:features:msg_comp}
    \end{subfigure}
    \hfill
    \begin{subfigure}[t]{0.33\textwidth}
        \centering
        \includegraphics[width=\textwidth]{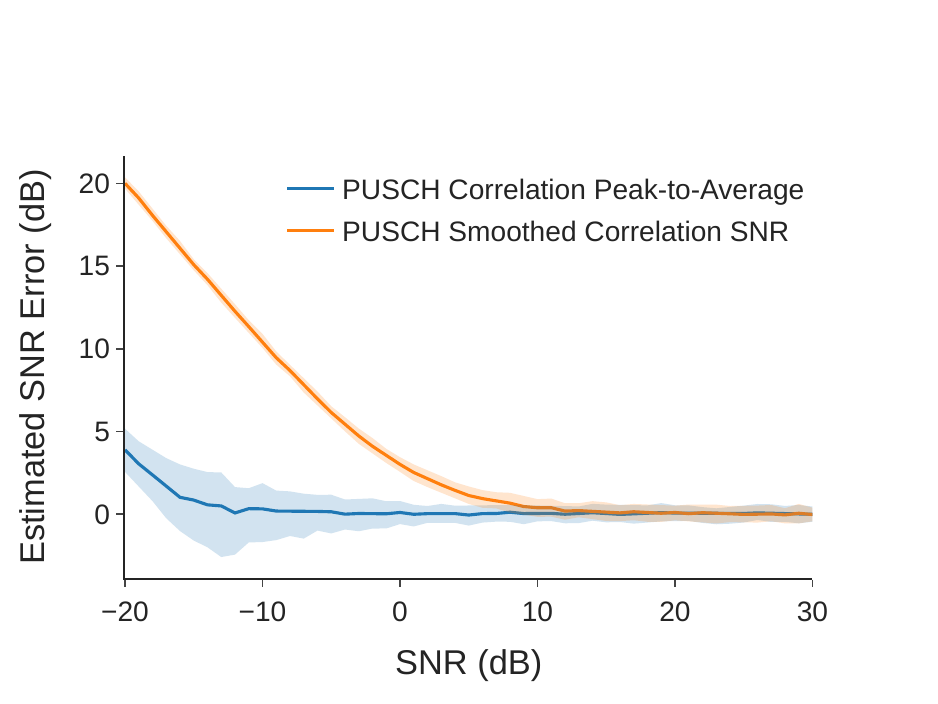} %
        \caption{\parbox{0.9\linewidth}{Error of SNR estimation with two different techniques.}}
        \label{fig:features:snr}
    \end{subfigure}
    
    \caption{Simulations of underlying features of our detection model. We varied SNR of the signal using AWGN channel. For each SNR, we ran the simulation 100 times.}
    \label{fig:features}
\end{figure*}

\section{Detailed Features Description}\label{sec:features_appendix}

As uplink signal features, we collect various signal power and signal quality features derived from the circular cross-correlation. The limiting factor in acquiring reference signals is the sampling rate of the signal, which translates into the same resolution of the cross-correlation. To increase this resolution, we upsample the resulting time domain cross-correlation and therefore infer information at a higher sampling rate~\cite{spacey_is_2012}. We use the Fourier method~\cite{noauthor_resample_nodate} for upsampling. 

\subsubsection{Signal Power}

A na\"{i}ve way to estimate the signal power is through RMS$^2$ signal power measurement of the corresponding PRBs. However, it is influenced by unrelated signals on the same frequencies, such as uplink traffic from other UEs on other cells or other interference. 

Instead, we use the interpolated correlation peak as an improved signal power measure. In \autoref{fig:features:sig_power}, we ran 100 simulations for each SNR value to compare the interpolated correlation peak to the classical RMS$^2$. Especially in low SNR conditions, it outperforms na\"{i}ve RMS$^2$ signal power calculation as seen in \autoref{fig:features:sig_power}. The correlation focuses on the target signal only and remains largely unaffected by noise, resulting in a more robust feature. In another simulation, we saw that the effect of a time delay on the signal power features is small and even with a time delay of 5$\mu$s, the error magnitude is only around 0.5 dB.

\subsubsection{Comparison of Uplink Message Types}

The measurement error of the features depends on the message type and its size. As shown in \autoref{fig:features:msg_comp}, the estimated signal power value based on correlation peaks is more accurate for PRACH and PUSCH with more PRBs than for PUCCH and PUSCH with only 1 PRB. This is because the PRACH includes the longest reference signal, allowing for more accurate correlation computations. In contrast, a PUSCH transmission with 1 PRB has only 2 reference signal sequences, each just 12 samples long, leading to poorer performance. A PUCCH transmission, despite spanning only 1 PRB, contains reference signals in 6 OFDM symbols, resulting in better performance than PUSCH with 1 PRB.

\subsubsection{Signal Quality}

To estimate signal quality, we use a more robust measurement technique compared to previous uplink sniffers \cite{kotuliak_ltrack_2022, hoang_ltesniffer_2023}. In their work, the circular cross-correlation is smoothed with a filter in the frequency domain to eliminate the impact of noise. The difference between the smoothed and non-smoothed version of the correlation is their resulting noise estimate. We call this technique \textit{Smoothed Correlation SNR Estimation}. For our estimation technique, called \textit{Correlation Peak-to-Average} measurement, we divide the peak value of the cross-correlation with the average value of the cross-correlation. The intuition is that the signal gets concentrated in and around the peak, while the noise is spread over the whole correlation. The comparison of the two features can be found in \autoref{fig:features:snr}.

In 5G OFDM symbols allocated to the PUSCH reference signals, some subcarriers are left unoccupied. We can thus measure noise in the band from these empty subcarriers ($N =$ RMS$^2$ of empty subcarriers). On the other hand, the signal at the occupied subcarrier is the relevant signal plus the noise ($P = S + N=$ RMS$^2$ of occupied subcarriers). Then the SNR for these OFDM symbols can be calculated as $P/N - 1$.

\section{Robustness Test}\label{sec:robust}

\begin{figure}[ht]
    \centering
    \includegraphics[width=1\linewidth]{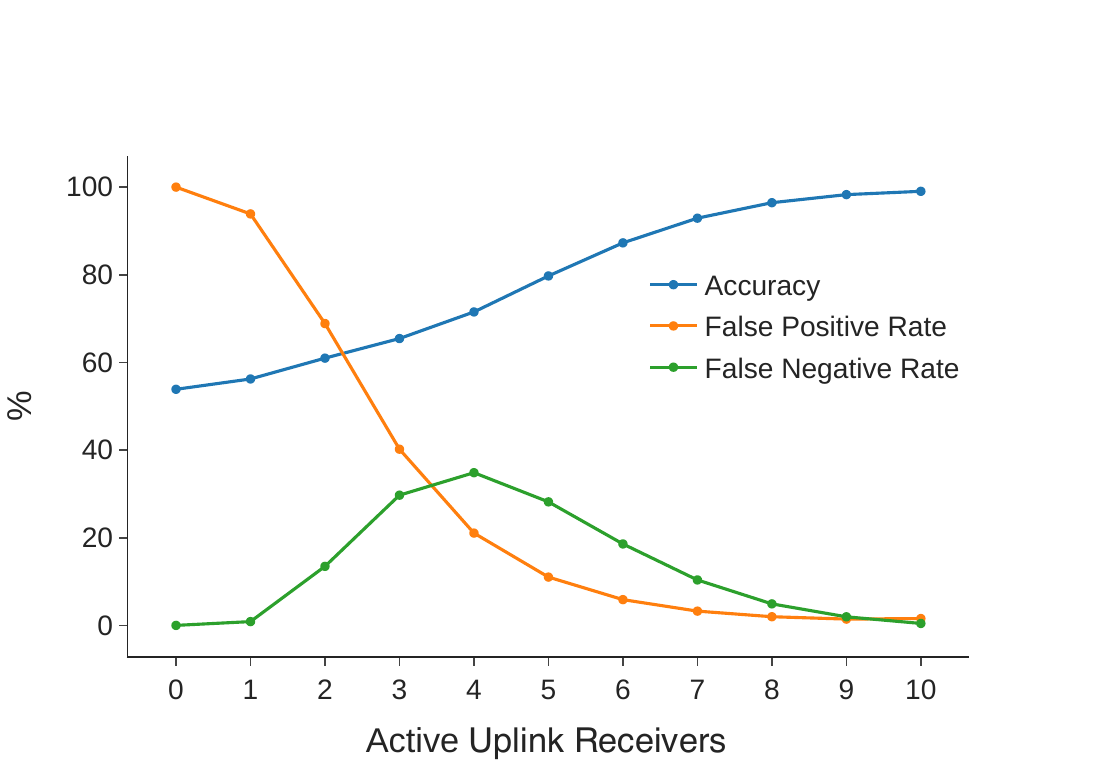}
    \caption{Effect of uplink receivers being switched off.}
    \label{fig:lru_going_down}
\end{figure}

We performed robustness tests to determine how well the model behaves if fewer uplink receivers are available or if fewer messages are available for classification. For the first test, we plot these results in \autoref{fig:lru_going_down}. In the second test, with just one message (PRACH), the model's accuracy is 99.17\%, and with the first two messages (PRACH and PUSCH (RRC Connection Request)), the model's accuracy improves to 99.6\%.

\end{appendices}

\end{document}